\documentclass[preprint2]{emulateapj}
\usepackage{graphicx}
\usepackage{txfonts}
\usepackage{url}
\graphicspath{{./fig/}{./png/}}
\usepackage{bm}

\newcommand{\EQ}{\begin{equation}}
\newcommand{\EE}{\end{equation}}
\newcommand{\EQA}{\begin{eqnarray}}
\newcommand{\EEA}{\end{eqnarray}}

\newcommand{\pd}{\partial}
\newcommand{\DIV}{\bm{\nabla} \cdot }

\newcommand{\mean}[1]{\overline{#1}}
\newcommand{\meanv}[1]{\overline{\bm #1}}

\newcommand{\cs}{c_{\rm s}}

\newcommand{\nut}{\nu_{\rm t}}
\newcommand{\nutz}{\nu_{\rm t0}}
\newcommand{\urms}{u_{\rm rms}}

\newcommand{\kef}{k_{\rm f}}

\newcommand{\rij}{R_{ij}}
\newcommand{\rxx}{R_{xx}}
\newcommand{\ryy}{R_{yy}}

\newcommand{\rxy}{R_{xy}}
\newcommand{\rxz}{R_{xz}}
\newcommand{\ryz}{R_{yz}}
\newcommand{\St}{{\rm St}}
\newcommand{\Sh}{{\rm Sh}}
\newcommand{\Ma}{{\rm Ma}}

\newcommand{\Pra}{{\rm Pr}}
\newcommand{\Ra}{{\rm Ra}}
\newcommand{\Rey}{{\rm Re}}

\newcommand{\Co}{{\rm Co}}

\def\onehalf{{\textstyle{1\over2}}}

\def\onethird{{\textstyle{1\over3}}}

\begin{document}

\title{Angular momentum transport in convectively unstable shear flows}

\author{Petri J. K\"apyl\"a\altaffilmark{1,2}, Axel Brandenburg\altaffilmark{2,3},
Maarit J. Korpi\altaffilmark{1}, Jan E. Snellman\altaffilmark{1}, and Ramesh Narayan\altaffilmark{4}}

\altaffiltext{1}{
Department of Physics, Division of Geophysics and Astronomy, FI-00014 University of Helsinki, Finland
}\altaffiltext{2}{
NORDITA, AlbaNova University Center, Roslagstullsbacken 23, SE-10691 Stockholm, Sweden
}\altaffiltext{3}{
Department of Astronomy, Stockholm University, SE-10691 Stockholm, Sweden
}\altaffiltext{4}{
Harvard-Smithsonian Center for Astrophysics
60 Garden Street, MS-51 Cambridge, MA 02138, USA
\\ $ $Revision: 1.143 $ $ (\today)
}

\begin{abstract}
  Angular momentum transport owing to hydrodynamic turbulent convection is
  studied using local three dimensional numerical simulations
  employing the shearing box approximation. 
  We determine the turbulent viscosity from non-rotating runs over
  a range of values of the shear parameter and
  use a simple analytical model in order
  to extract the non-diffusive contribution
  ($\Lambda$-effect) to the stress in runs where rotation is 
  included.
  Our results suggest that the turbulent viscosity is of the order of
  the mixing length estimate and weakly affected
  by rotation. The $\Lambda$-effect is non-zero and a
  factor of 2--4 smaller than the turbulent viscosity in the slow
  rotation regime.
  We demonstrate that for Keplerian shear,
  the
  angular momentum transport can change sign and be outward
  when the rotation period is greater than the turnover time,
  i.e.\ when the Coriolis number is below unity.
  This result seems to be relatively independent of the value of
  the Rayleigh number.
\end{abstract}

\keywords{accretion, accretion disks -- convection -- stars: rotation
  -- Sun: rotation -- turbulence }

\section{Introduction}
Turbulence due to the convective instability is thought to account for
much of the angular momentum transport in the outer layers of the Sun
and other stars with convection zones (e.g.\ R\"udiger 1989;
R\"udiger \& Hollerbach 2004). In the
presence of turbulence the fluid mixes efficiently and diffusion
processes occur much faster than in its absence. This
effect is usually parameterized by a turbulent viscosity $\nut$ that
is much larger than the molecular viscosity $\nu$. Often the value of
$\nut$ is estimated using simple mixing length arguments with
$\nut=\urms l/3$, where $\urms$ is the rms velocity of the turbulence
and $l=\alpha_{\rm MLT} H$ where $\alpha_{\rm MLT}$ is a parameter of
the order unity and $H$ is the vertical pressure scale height.
Numerical results from
simpler fully periodic isotropically forced systems suggest that the
mixing length estimate gives the correct order of magnitude of
turbulent viscosity (e.g.\ Yousef et al.\ 2003;
K\"apyl\"a et al.\ 2009a; Snellman et al.\
2009). However, it is important to compute $\nut$ from convection
simulations in order to see whether the results of the simpler systems
carry over to convection. Furthermore, it is of interest to study
whether the small-scale turbulent transport can be understood in the light
of simple analytical closure models that can be used in
subgrid-scale modeling. Measuring $\nut$ and its relation to averaged
quantities, such as correlations of turbulent velocities, is one of 
the main purposes of our study.

In addition to enhanced viscosity, turbulence can also lead to
non-diffusive transport. The $\alpha$-effect (e.g.\ Krause \&
R\"adler 1980), responsible for the generation of large-scale magnetic
fields by helical turbulence, is one of the most well-known
non-diffusive effects of turbulence. An analogous effect exists in
the hydrodynamical regime and is known as the $\Lambda$-effect (Krause \&
R\"udiger 1974). The $\Lambda$-effect is proportional to the local
angular velocity and occurs if the turbulence is anisotropic in
the plane perpendicular to the rotation vector (R\"udiger 1989). The
existence of the $\Lambda$-effect has been established numerically
from convection simulations (e.g.\ Pulkkinen et al.\ 1993; Chan 2001; K\"apyl\"a
et al.\ 2004; R\"udiger et al.\ 2005) and simpler homogeneous systems
(K\"apyl\"a \& Brandenburg 2008).

If, however, both shear and rotation are present it is difficult to
disentangle the diffusive and non-diffusive contributions. This is
particularly important in the case of accretion disks where the sign
of the stress determines whether angular momentum is transported
inward or outward.
Convection is commonly not considered as a viable angular momentum transport
mechanism in accretion disks since several studies have indicated that
the transport owing to convection occurs inward
(e.g.\ Cabot \& Pollack 1992; Ryu \& Goodman 1992; Stone \& Balbus 
1996; Cabot 1996; R\"udiger et al.\ 2002). Furthermore, in an influential
paper, Stone \& Balbus (1996, hereafter SB96) presented numerical
simulations of hydrodynamic convection where the transport was
indeed found to be small and directed inward on average.
This result was used to provide additional evidence for the importance
of the magneto-rotational instability (Balbus \& Hawley 1991)
as the main mechanism providing angular momentum transport in
accretion disks.
Although we agree with the conclusion that hydrodynamic turbulence
is ineffective in providing angular momentum transport,
we are concerned about the generality of the result of SB96.
There are now some indications that hydrodynamic turbulence
may not always transport angular momentum inward (cf.\ Lesur
\& Ogilvie 2010).

In order to approach the problem from a more general perspective,
Snellman et al.\ (2009) studied isotropically
forced turbulence under the influence of shear and rotation and found
that the total stress, corresponding to the radial angular momentum
transport in an accretion disk, can change sign as rotation and
shear of the system are varied in such a way that their ratio remains
constant. They found that the stress is positive for small Coriolis
numbers, corresponding to slow rotation.
In what follows we show that outward transport is possible also for
convection in a certain range of Coriolis numbers.
When rotation is slow, the Reynolds stress is
positive, corresponding to outward transport. In the regime of large Coriolis
numbers the Reynolds stress changes sign and is directed inward, as in the
study of SB96.

The remainder of the paper is organized as follows: our numerical
model is described in Section~\ref{sec:model} and the relevant
mean-field description in Section~\ref{sec:mf}. The results and the
conclusions are given in Sections~\ref{sec:results} and \ref{sec:conc},
respectively.

\section{Model}
\label{sec:model}

Our model is the same as that used in K\"apyl\"a et al.\ (2008),
but without magnetic fields. 
We solve the following set of equations for compressible hydrodynamics,
\begin{equation}
\frac{\mathcal{D} \ln \rho}{\mathcal{D}t} = -\DIV{\bm U},
\end{equation}
\begin{equation}
 \frac{\mathcal{D} \bm U}{\mathcal{D}t} = -SU_x\bm{\hat{y}}-\frac{1}{\rho}{\bm \nabla}p + {\bm g} - 2\bm{\Omega} \times \bm{U} + \frac{1}{\rho} \bm{\nabla} \cdot 2 \nu \rho \mbox{\boldmath ${\sf S}$}, \label{equ:UU}
\end{equation}
\begin{equation}
 \frac{\mathcal{D} e}{\mathcal{D}t} = - \frac{p}{\rho}\DIV {\bm U} + \frac{1}{\rho} \bm{\nabla} \cdot K \bm{\nabla}T + 2 \nu \mbox{\boldmath ${\sf S}$}^2 - Q, \label{equ:ene}
\end{equation}
where $\mathcal{D}/\mathcal{D}t = \pd/\pd t + (\bm{U} + \meanv{U}_0)
\cdot \bm{\nabla}$ is the total advective derivative,
$\meanv{U}_0 = (0,Sx,0)$ is the imposed
large-scale shear flow, $\nu$ is the
kinematic viscosity, $K$ is the
heat conductivity, $\rho$ is the density, $\bm{U}$ is the velocity, $\bm{g}$
is the gravitational acceleration, and
$\bm{\Omega}=\Omega_0\hat{\bm{z}}$ is the rotation vector. The fluid
obeys an ideal gas law $p=\rho e (\gamma-1)$, where $p$ and $e$ are
pressure and internal energy, respectively, and $\gamma = c_{\rm
  P}/c_{\rm V} = 5/3$ is the ratio of specific heats at constant
pressure and volume, respectively. The specific internal energy per
unit mass is related to the temperature via $e=c_{\rm V} T$. The
traceless rate-of-strain tensor $\mbox{\boldmath ${\sf S}$}$ is given
by
\begin{equation}
{\sf S}_{ij} = \onehalf (U_{i,j}+U_{j,i}) - \onethird \delta_{ij} \DIV \bm{U},
\end{equation}
where commas denote partial differentiation.

We use a three layer, piecewise polytropic stratification 
with constant gravity ${\bm g}=-g\bm{\hat{z}}$. 
The vertical variation of $g$ in real accretion disks is neglected
in our local model.
The positions of the bottom of the box, bottom and top
of the convectively unstable layer, and the top of the box,
respectively, are given by $(z_1, z_2, z_3, z_4) = (-0.85, 0, 1,
1.15)d$, where $d$ is the depth of the convectively unstable
layer. Initially the stratification is piecewise polytropic with
polytropic indices $(m_1, m_2, m_3) = (3, 1, 1)$. The horizontal
extent of the box is twice the vertical extent, i.e.\
$L_x/d=L_y/d=2L_z/d=4$. The thermal conductivity has a vertical
profile that maintains a convectively unstable layer above a stable
layer at the bottom of the domain. A prescribed temperature gradient
at the base maintains a constant heat flux into the domain.
The last term in Equation~(\ref{equ:ene}) describes an externally applied
cooling with
\begin{equation}
Q=\frac{e-e_0}{\tau_{\rm cool}(z)},
\end{equation}
where $e_0$ is the internal energy at $z_4$ and $\tau_{\rm cool}(z)$ is a cooling
time which is constant for $z>z_3$ and smoothly connects to the
lower layer where $\tau_{\rm cool}\rightarrow\infty$.

We use impenetrable stress-free boundary conditions at the
top and bottom boundaries for the velocity
\begin{equation}
\frac{\pd U_x}{\pd z}=\frac{\pd U_y}{\pd z}=U_z=0.
\end{equation}
The temperature gradient at the bottom of the domain is given by
\begin{equation}
\frac{\pd T}{\pd z} = \frac{-g/c_{\rm V}}{(m_1+1)(\gamma-1)},
\end{equation}
where $m_1=3$ is the polytropic index at $z_1$. All quantities are
periodic in the $y$-direction, whereas shearing periodic conditions
(Wisdom \& Tremaine 1988) are used in the $x$-direction.
The same setup has been used in earlier work to model convection in
local patches in a star, but here we apply it also to a layer near the
surface of an accretion disk.
The source of heating in the midplane of the disk is not specified
and is instead assumed given.
This is appropriate for addressing the more general question
about the direction of angular momentum transport once there is
convection in the absence of a magnetic field.

\begin{figure*}
\centering
\includegraphics[width=1.\textwidth]{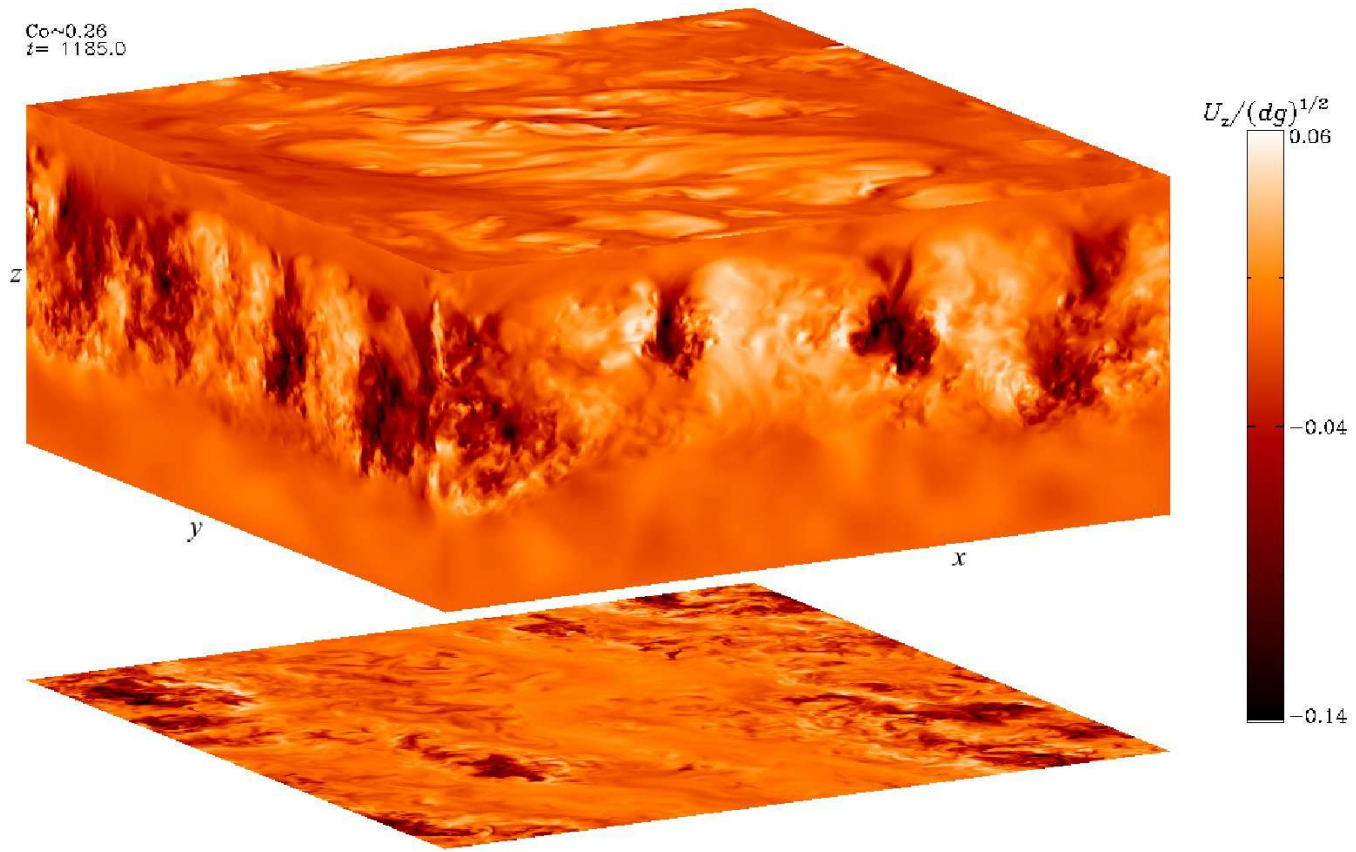}
\caption{Snapshot of the vertical velocity from run~D5 with
  $\Rey\approx648$. The sides of the box show the periphery of the
  domain whereas the top and bottom slices show $U_z$ at vertical
  heights $z = 0.95d$ and $z = 0$, respectively. Here, $\Co=0.26$ and 
  $\Sh=-0.20$. See also \texttt{http://www.helsinki.fi/\ensuremath{\sim}kapyla/movies.html}
  for an animation.
}
   \label{fig:snap}
\end{figure*}

We employ the {\sc Pencil Code}\footnote{http://pencil-code.googlecode.com/},
which is a high-order finite difference code for solving the compressible
MHD-equations.
The bulk of our simulations were performed at a moderate resolution of
$128^3$ grid points. In a few cases we study the behavior of the
solutions at higher resolutions (up to $1024^3$), see
Figure~\ref{fig:snap} for a snapshot of a high resolution simulation,
showing the vertical velocity on the periphery of the domain.

\subsection{Dimensionless quantities and parameters}

Non-dimensional quantities are derived such that
\begin{eqnarray}
d = g = \rho_0 = c_{\rm P} = 1,
\end{eqnarray}
where $\rho_0$ is the initial density at $z_2$. The units of length, time,
velocity, density, and entropy are
\begin{eqnarray}
[x] = d,\; [t] = \sqrt{d/g},\; [U]=\sqrt{dg},\; [\rho]=\rho_0,\; [s]=c_{\rm P}.
\end{eqnarray}
The system is characterized by several non-dimensional parameters.
We define the Prandtl number and the Rayleigh number as
\begin{eqnarray}
\Pra=\frac{\nu}{\chi_0},\;\; \Ra=\frac{gd^4}{\nu \chi_0} \bigg(-\frac{1}{c_{\rm P}}\frac{{\rm d}s}{{\rm d}z
} \bigg)_0,
\end{eqnarray}
where $\chi_0 = K/(\rho_{\rm m} c_{\rm P})$ is a reference value for the thermal
diffusivity, and $\rho_{\rm m}$ is the density at the mid layer $z_{\rm
  m}=\onehalf(z_3-z_2)$. The entropy gradient, measured at $z_{\rm m}$
in the non-convecting hydrostatic state, is given by
\begin{eqnarray}
\bigg(-\frac{1}{c_{\rm P}}\frac{{\rm d}s}{{\rm d}z}\bigg)_0 = \frac{\nabla-\nabla_{\rm ad}}{H_{\rm P}},
\end{eqnarray}
where $\nabla-\nabla_{\rm ad}$ is the superadiabatic temperature
gradient with $\nabla_{\rm ad} = 1-1/\gamma$, $\nabla = (\pd \ln T/\pd
\ln p)_{z_{\rm m}}$, and $H_{\rm P}$ is the pressure scale height
at $z_{\rm m}$.
The amount of stratification is determined by the parameter $\xi_0
=(\gamma-1) e_0/(gd)$, which is the pressure scale height at the top
of the domain normalized by the depth of the unstable layer. We use in
all cases $\xi_0 ={1\over3}$, which results in a total density contrast of
about 23.
The Mach number in our simulations is of the order of $0.1$ or less.

We define the Coriolis and shear numbers, describing the
strengths of rotation and shear, respectively, as
\begin{eqnarray}
\Co = \frac{2\Omega_0}{\urms \kef}, \;\;\; \Sh = \frac{S}{\urms \kef},
\label{Codef}
\end{eqnarray}
where $\urms$ is the rms value of the turbulent velocity
averaged over the full volume,
$\kef=2\pi/d$ is an estimate of the energy carrying wavenumber, and $d$
is the depth of the convectively unstable layer. The Reynolds
number is given by
\begin{eqnarray}
{\rm Re} = \frac{\urms}{\nu \kef}.
\end{eqnarray}
For accretion disk applications it is convenient to define
also the relative shear rate,
\begin{equation}
q=-\frac{S}{\Omega_0},
\end{equation}
which describes the rotation profile of the disk, where the
local angular velocity varies like $\Omega \propto r^{-q}$.
For Keplerian disks we have $q=3/2$.

The majority of the simulations ran for $10^3$ time units
($\Delta t=10^3\sqrt{d/g}$), which
corresponds to roughly 250 convective turnover times; $\tau_{\rm turn}=(\urms
\kef)^{-1}$. For the analysis the first 50 turnover times of the simulations were
usually discarded in order to minimize the effects of initial
transients. The highest resolution runs with $512^3$ and $1024^3$ grid
points were started by remeshing from snapshots of lower resolution runs and ran
for 60 and 14 turnover times, respectively.

Error bars are estimated by dividing the time series into three equally
long parts and computing averages for each part individually. The
largest departure from the average over the full time series is taken
to represent the error.

\begin{deluxetable}{ccccccccc}
\tabletypesize{\scriptsize}
\tablecaption{Summary of the Runs with Variable $q$. Here, $\mbox{\rm
    Ma}=\urms/(gd)^{1/2}$ and $\tilde{R}_{xy}=\langle\rxy\rangle/\urms^2$, 
    where $\langle \rxy \rangle$ is the volume average of the stress over
    the convectively unstable layer.
    $\Pra=1$ and $\Ra=10^6$ in all runs.}
\tablewidth{0pt}
\tablehead{
\colhead{Run} & \colhead{$q$} & \colhead{$\Co$} & \colhead{$\Sh$} & 
\colhead{$\Rey$} & 
\colhead{${\rm Ma}$} & \colhead{$\tilde{R}_{xy}$} & \colhead{grid}
}
\startdata
A0    &    -     &    0    & $ 0.00$ & 28  & 0.036 & $0.003\pm0.003$ & $128^3$ \\ 
A1    &    -     &    0    & $-0.04$ & 30  & 0.037 & $0.064\pm0.020$ & $128^3$ \\ 
A2    &    -     &    0    & $-0.08$ & 31  & 0.039 & $0.110\pm0.016$ & $128^3$ \\ 
A3    &    -     &    0    & $-0.14$ & 33  & 0.041 & $0.152\pm0.040$ & $128^3$ \\ 
A4    &    -     &    0    & $-0.17$ & 37  & 0.047 & $0.172\pm0.040$ & $128^3$ \\ 
\hline
B1    & $-50.00$ & $-0.01$ & $-0.19$ & 33  & 0.042 & $0.149\pm0.022$ & $128^3$ \\ 
B2    & $-25.00$ & $-0.02$ & $-0.19$ & 34  & 0.042 & $0.151\pm0.014$ & $128^3$ \\ 
B3    & $-10.00$ & $-0.04$ & $-0.20$ & 32  & 0.040 & $0.176\pm0.017$ & $128^3$ \\ 
B4    &  $-5.00$ & $-0.09$ & $-0.22$ & 29  & 0.036 & $0.147\pm0.012$ & $128^3$ \\ 
B5    &  $-2.50$ & $-0.18$ & $-0.23$ & 28  & 0.035 & $0.119\pm0.021$ & $128^3$ \\ 
B6    &  $-1.00$ & $-0.49$ & $-0.25$ & 26  & 0.032 & $0.069\pm0.010$ & $128^3$ \\ 
B7    &  $-0.50$ & $-1.02$ & $-0.25$ & 25  & 0.031 & $0.033\pm0.001$ & $128^3$ \\ 
B8    &  $-0.25$ & $-2.25$ & $-0.28$ & 23  & 0.028 & $0.014\pm0.002$ & $128^3$ \\ 
B9    &  $-0.10$ & $-7.67$ & $-0.38$ & 17  & 0.021 & $0.015\pm0.005$ & $128^3$ \\ 
\hline
C1    &  $1.99$ &  $0.20$  & $-0.20$ & 32  & 0.040 & $0.051\pm0.013$ & $128^3$ \\ 
C2    &  $1.75$ &  $0.24$  & $-0.21$ & 30  & 0.038 & $0.039\pm0.024$ & $128^3$ \\ 
C3    &  $1.50$ &  $0.30$  & $-0.22$ & 28  & 0.036 & $0.053\pm0.003$ & $128^3$ \\ 
C4    &  $1.25$ &  $0.38$  & $-0.24$ & 26  & 0.033 & $0.076\pm0.014$ & $128^3$ \\ 
C5    &  $1.00$ &  $0.49$  & $-0.25$ & 26  & 0.032 & $0.066\pm0.006$ & $128^3$ \\ 
C6    &  $0.75$ &  $0.67$  & $-0.25$ & 25  & 0.032 & $0.047\pm0.007$ & $128^3$ \\ 
C7    &  $0.50$ &  $1.02$  & $-0.25$ & 25  & 0.031 & $0.014\pm0.006$ & $128^3$ \\ 
C8    &  $0.25$ &  $2.19$  & $-0.27$ & 23  & 0.029 & $-0.015\pm0.005$ & $128^3$ \\ 
C9    &  $0.10$ &  $7.46$  & $-0.37$ & 17  & 0.021 & $-0.036\pm0.001$ & $128^3$    
\enddata
\label{tab:runs}
\end{deluxetable}

\begin{deluxetable}{cccccccccc}
\tabletypesize{\scriptsize}
\tablecaption{Summary of the Runs with Keplerian Shear, i.e.\ here
  $q=1.5$ in all Runs.}
\tablewidth{0pt}
\tablehead{
\colhead{Run} & \colhead{$\Pra$} & \colhead{$\Ra$} & 
\colhead{$\Co$} & \colhead{$\Sh$} & \colhead{$\Rey$} & 
\colhead{${\rm Ma}$} & \colhead{$\tilde{R}_{xy}$} & \colhead{grid}
}
\startdata
D1    &            1 &        $10^6$  & 0.30 & $-0.22$ &  28 & 0.036 & $0.053\pm0.003$ &  $128^3$ \\ 
D2    &  ${1\over2}$ &  $2\cdot10^6$  & 0.29 & $-0.21$ &  59 & 0.037 & $0.053\pm0.028$ &  $256^3$ \\ 
D3    &  ${1\over4}$ &  $4\cdot10^6$  & 0.27 & $-0.20$ & 125 & 0.039 & $0.033\pm0.004$ &  $256^3$ \\ 
D4    & ${1\over10}$ &        $10^7$  & 0.27 & $-0.20$ & 315 & 0.039 & $0.018\pm0.006$ &  $512^3$ \\ 
D5    & ${1\over20}$ &  $2\cdot10^7$  & 0.26 & $-0.20$ & 648 & 0.041 & $0.031\pm0.010$ & $1024^3$ \\ 
\hline
E1    & 1 & $10^6$ & $0.01$ & $-0.01$ & 27  & 0.034 & $0.022\pm0.021$ & $128^3$ \\ 
E2    & 1 & $10^6$ & $0.03$ & $-0.02$ & 26  & 0.033 & $0.032\pm0.022$ & $128^3$ \\ 
E3    & 1 & $10^6$ & $0.06$ & $-0.05$ & 28  & 0.035 & $0.056\pm0.013$ & $128^3$ \\ 
E4    & 1 & $10^6$ & $0.12$ & $-0.09$ & 27  & 0.034 & $0.084\pm0.007$ & $128^3$ \\ 
E5    & 1 & $10^6$ & $0.29$ & $-0.21$ & 29  & 0.037 & $0.045\pm0.013$ & $128^3$ \\ 
E6    & 1 & $10^6$ & $0.63$ & $-0.47$ & 27  & 0.034 & $0.074\pm0.011$ & $128^3$ \\ 
E7    & 1 & $10^6$ & $1.32$ & $-0.99$ & 25  & 0.032 & $0.041\pm0.011$ & $128^3$ \\ 
E8    & 1 & $10^6$ & $4.33$ & $-3.24$ & 20  & 0.025 & $-0.040\pm0.003$ & $128^3$  
\enddata
\label{tab:kepruns}
\end{deluxetable}

\section{Mean-field interpretation}
\label{sec:mf}
In mean-field hydrodynamics the velocity field is decomposed
into its mean and fluctuating parts,
\begin{equation}
\bm{U}=\mean{\bm U}+\bm{u},\label{equ:udeomp}
\end{equation}
where the overbar denotes averaging and lowercase $\bm{u}$ the
fluctuation. In the present paper we consider horizontal averaging so
that the mean quantities depend only on $z$. We define the Reynolds
stress as
\begin{equation}
R_{ij}=\mean{u_i u_j}.
\end{equation}
Fluctuations of the density are here ignored for simplicity.

The Reynolds stress is
often described in terms of the Boussinesq ansatz which relates the
stress to the symmetrized gradient matrix of the large-scale velocity
\begin{equation}
R_{ij} = -\mathcal{N}_{ijkl}\mean{U}_{k,l}+\ldots,\label{equ:Ban}
\end{equation}
where the dots indicate higher derivatives of $\meanv{U}$ that
can occur in the expansion. The expression (\ref{equ:Ban}) states that
the stress is diffusive in character. In the general case the
fourth rank tensor $\mathcal{N}_{ijkl}$ can have a complicated structure,
see e.g.\ R\"udiger (1989). However, if we assume that the shear is
weak, the simplest description of the horizontal stress generated by our linear
shear flow is given by
\begin{equation}
R_{xy}=-\nut (\mean{U}_{x,y}+\mean{U}_{y,x})
=-2\nut{\sf S}_{xy}=-\nut S,\label{equ:nut}
\end{equation}
where $\nut=\nut(z)$ is the $z$-dependent turbulent viscosity, and
the component ${\sf S}_{xy}$ of the rate-of-strain tensor is not
to be confused with the shear rate $S$.
To our knowledge, SB96 present the only published results of turbulent
viscosity in the absence of rotation as
determined from convection simulations with large-scale shear, and 
they only provide a
volume averaged quantity for one case.
If also rotation is present (cf.\ Cabot 1996; Lesur \& Ogilvie 2010)
the Reynolds stress can no longer be related to turbulent viscosity
alone (see below).
In the present paper we study the dependence between stress and shear
systematically and estimate the turbulent viscosity coefficient
$\nut$.

It turns out that in many applications, Equation~(\ref{equ:Ban}) is
insufficient to describe the stress.
For example, according to Equation~(\ref{equ:nut}), the Reynolds stress
component $R_{\theta \phi}$ derived from observations of sunspot proper motions with the
observed surface differential rotation would yield $\nut<0$,
which is clearly unphysical (see the
discussions in Tuominen \& R\"udiger 1989; and Pulkkinen et al.\
1993).
This motivates the inclusion of a non-diffusive contribution proportional
to the rotation of the system (e.g., Wasiutynski 1946), such that
\begin{equation}
R_{ij} = \Lambda_{ijk}\Omega_k -\mathcal{N}_{ijkl}\mean{U}_{k,l}+\ldots,\label{equ:totstress}
\end{equation}
where $\Lambda_{ijk}$ are the components of the $\Lambda$-effect.
This effect is expected to occur in anisotropic
turbulence under the influence of rotation (e.g.\ R\"udiger 1989). In
convection the density stratification provides the
anisotropy.
This is confirmed by numerous simulations of rigidly rotating stratified
convection (e.g.\ Pulkkinen et al.\ 1993; Chan 2001;
K\"apyl\"a et al.\ 2004; R\"udiger et al. 2005). Although additional
shear flows are generated in these systems when the gravity and
rotation vectors are not parallel or antiparallel, no serious attempt
has been made to quantify the turbulent viscosity in convection.

When shear and rotation
are both present, it is no longer possible to distinguish between the
diffusive and non-diffusive contributions without resorting to
additional theoretical arguments. 
Here we use a simple
analytical approach based on the so-called minimal
$\tau$-approximation (see e.g.\ Blackman \& Field 2002, 2003)
to estimate the contributions from
$\nut$ and the 
$\Lambda$-effect.

The idea behind the minimal $\tau$-approximation is to use relaxation terms 
of the form $-\tau^{-1}R_{ij}$ in the evolution equations for the
components of the Reynolds stress $\rij$ as a simple description 
of the nonlinearities.
Using the decomposition~(\ref{equ:udeomp})
and the Navier-Stokes equations one can derive equations 
for the Reynolds stress. For the purposes of the present study 
it suffices 
to consider a situation with one spatial dimension, $z$. In this case the evolution 
equations can be written as
\begin{eqnarray}
\pd_t R_{ij} &=& -\mean{U}_z \pd_z R_{ij}-R_{iz} \pd_z \mean{U}_j-R_{jz} \pd_z \mean{U}_i -S \delta_{yi} R_{xj} \nonumber \\
& & -S \delta_{yj} R_{xi} - 2 \epsilon_{ilk}\Omega_l R_{kj} - 2 \epsilon_{jlk}\Omega_l R_{ki}+ N_{ij},
\label{equ:taulong}
\end{eqnarray}
where $N_{ij}$ represents nonlinear terms. 
As described above, using the
minimal $\tau$-approximation as closure model, one substitutes 
\begin{equation}
N_{ij}=-\tau^{-1}R_{ij},
\label{equ:taushort}
\end{equation}
where $\tau$ is a relaxation time, which is here treated as
a free parameter that we expect to be close to $\tau_{\rm turn}$.
In the present paper, however, we will not solve the closure model
self-consistently but rather compare the leading terms with the
numerical results. A more thorough investigation using the full closure model
is postponed to a later study.

\section{Results}
\label{sec:results}
In order to study the Reynolds stress generated by shear and
rotation, we perform five sets of simulations summarized in
Tables~\ref{tab:runs} and \ref{tab:kepruns}. 
Our base run is one with $\Rey\approx30$,
$\Pra=1$, and $\Co=\Sh=0$ (run~A0). In set~A we add only shear and in sets~B
and C we keep the shear constant and vary rotation so that
$q$ is negative and positive, respectively. 
In the remaining sets of calculations we study the
Keplerian shear case ($q=1.5$): in set~D we vary $\Rey$ and thus $\Ra$
with fixed shear and rotation whereas in set~E we vary both $\Co$ and
$\Sh$ with fixed $\Rey$.

\subsection{Only shear, $\Sh\neq0$, $\Co=0$}
\label{OnlyShear}
\begin{figure}
\centering
\includegraphics[width=1.\columnwidth]{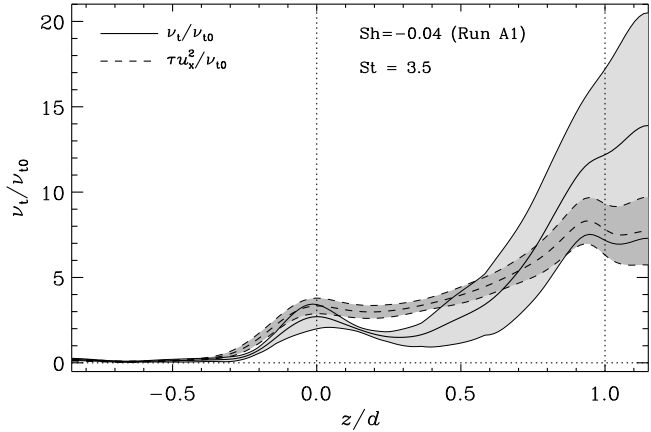}
\caption{Vertical profile of the
  turbulent viscosity $\nut$ (thick solid line), computed from 
  Equation~(\ref{equ:nutz}), and the analytical estimate (thick dashed 
  line) according to Equation~(\ref{equ:mtanut})
  for run~A1 with $\Sh=-0.04$, $\Co=0$ and $\Rey=30$. The shaded areas
  show the error estimates. The dotted vertical lines at $z=0$ and
  $z=d$ denote the bottom and top of the convectively unstable layer,
  respectively.}
   \label{fig:nut}
\end{figure}

\begin{figure}
\centering
\includegraphics[width=1.\columnwidth]{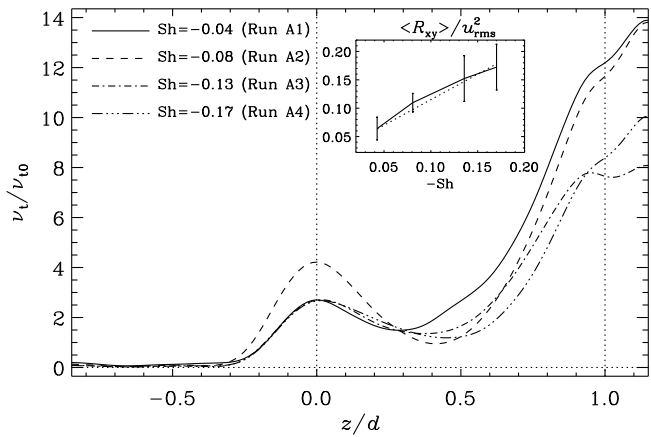}
\caption{Turbulent viscosity $\nut$ as a function of height
  for different shear parameters (runs~A1--A4). The inset shows $\rxy$
  averaged over the convectively unstable layer as a function of
  $\Sh$; dotted line is proportional to $\Sh$.}
   \label{fig:pnut}
\end{figure}

When only shear is present, the turbulent viscosity can be computed
from Equation~(\ref{equ:nut}) as
\begin{equation}
\nut=-R_{xy}^{(S)}/S,\label{equ:nutz}
\end{equation}
where the superscript $S$ indicates that the stress is due to the
large-scale shear. By virtue of density stratification, the
components of the stress tensor, and hence $\nut$, are functions of depth, i.e.\
$R_{ij}=R_{ij}(z)$ and $\nut=\nut(z)$.
We normalize our results with the estimate
\begin{equation}
\nutz=\onethird \tau \mean{\bm{u}^2}.
\label{nutz}
\end{equation}
Assuming that
the Strouhal number
\begin{equation}
\St=\tau\urms\kef,\label{equ_St}
\end{equation}
is equal to unity, i.e.\ $\tau=\tau_{\rm turn}$, we obtain
\begin{equation}
\nutz=\onethird \urms\kef^{-1}.\label{equ:nut0}
\end{equation}
Note that, if we allow $\St\neq1$, we have $\nutz=\onethird \St \urms\kef^{-1}$ 
and the ratio $\nut/\nutz$ gives a measure of the Strouhal number. 
A typical example of $\nut$ is shown
in Figure~\ref{fig:nut}. We find that the stress owing to the large-scale
shear is always positive, yielding $\nut>0$.
The stress $R_{xy}^{(S)}$
increases roughly proportionally to the shear parameter $S$ in the
range $0.04<|\Sh|<0.17$ so that the ratio $\nut/\nutz$ remains
approximately constant; see Figure~\ref{fig:pnut}. Increasing the
shear much beyond our relatively low maximum value leads to
large-scale vorticity generation. These flows often grow so strong
that the Mach number reaches unity leading to numerical difficulties.
We associate this phenomenon with the `vorticity dynamo' observed in
forced turbulence simulations (Yousef et al.\ 2008; K\"apyl\"a et al.\
2009a) and several theoretical studies (e.g.\ Elperin et al.\ 2003,
2007). 
Although the large-scale vorticity generation is weak in our 
runs~A1--A4 we see that in the absence of rotation even modest 
shear increases the rms-velocity measurably (see Table~\ref{tab:runs}).
However, when rotation is added, this instability vanishes 
(see, e.g., Snellman et al.\ 2009).

Our estimate for $\nutz$ is based on the volume averaged rms-velocity and a
somewhat arbitrarily defined length scale $1/\kef$. These choices are partly
responsible for the large values of $\nut/\nutz$. 
In order to obtain a more accurate
estimate we derive an evolution equation for $\rxy$ using 
Equation~(\ref{equ:taulong})
\begin{equation}
\pd_t\rxy=-\mean{U}_z\pd_z\rxy - \ryz \pd_z \mean{U}_x - \rxz \pd_z \mean{U}_y  -\mean{u_x^2}S + N_{xy}.
\end{equation}
In general the first term on the rhs is non-zero in the compressible
case but, as our Mach numbers are small, $\mean{U}_z$ is
negligible. Furthermore, the stress components $\rxz$ and $\ryz$
vanish under the assumption that the imposed shear is the only
large-scale velocity component that depends on the horizontal
coordinates. Thus the only terms remaining are
\begin{equation}
\pd_t\rxy=-\mean{u_x^2}S + N_{xy}.
\end{equation}
We now ignore the time derivative and apply the minimal
$\tau$-approximation, i.e.\ Equation~(\ref{equ:taushort}), to obtain
\begin{equation}
\rxy=-\tau\mean{u_x^2}S=-\nut S.\label{equ:mtanut}
\end{equation}
Note that we use $\tau$ as a fitting parameter when comparing 
the different
sides of the equation.
We find that the simple analytical estimate can be fitted with the
stress from the simulations when the Strouhal number is in the range
$3\ldots4$ for all runs in set~A, e.g., see the comparison shown in
Figure~\ref{fig:nut} for run~A1.
Here, for simplicity, we have assumed that $\tau$ has no dependence
on $z$ which can contribute to the fact that the curves have somewhat
different depth dependencies. Strouhal numbers in the range 1--3
are in line with previous numerical findings with forced
turbulence (e.g.\ Brandenburg et al.\ 2004) and convection (e.g.\
K\"apyl\"a et al.\ 2009b).

\subsection{Shear and rotation, $\Sh, \Co\neq0$}
\begin{figure}
\centering
\includegraphics[width=1.\columnwidth]{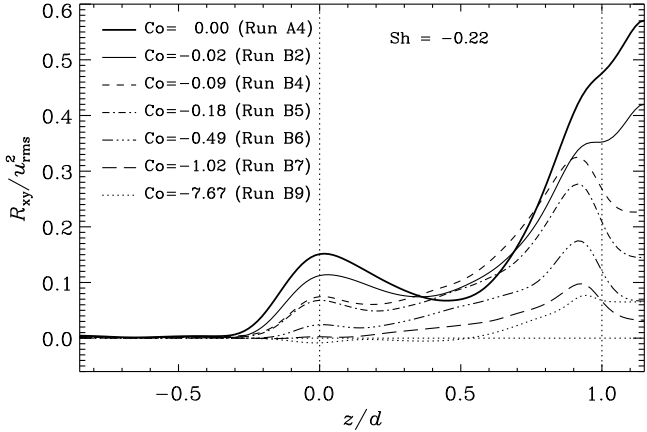}
\caption{Total stress $\rxy$ for different rotation rates using
a subset of runs in set~B with negative $q$,
  $\Sh\approx-0.22$, and $\Rey=30$.}
   \label{fig:pstressCon}
\end{figure}

\begin{figure}
\centering
\includegraphics[width=1.\columnwidth]{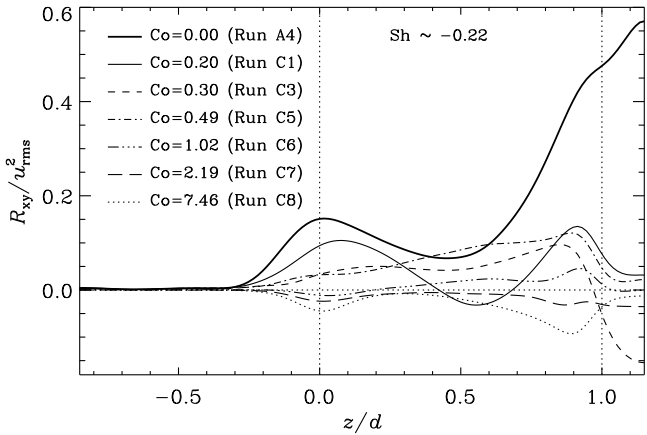}
\caption{Same as Figure~\ref{fig:pstressCon}, but for a subset of runs
  from set~C with positive values of $q$ (see the legend).}
   \label{fig:pstressCop}
\end{figure}

Figures~\ref{fig:pstressCon} and \ref{fig:pstressCop} show the total
stress $\rxy$ from sets~B and C where the imposed shear with 
$S=-0.05\sqrt{g/d}$ is kept constant
and rotation is varied in a way that $q$ is either negative (set~B) or
positive (set~C), respectively. 
For shear parameters $q>2$, the flow is Rayleigh unstable; thus,
we investigate the parameter regime from slightly below 2 down to
larger negative values.
Note that although the imposed shear is constant, the value of $\Sh$ varies
somewhat because it is based on the turbulent velocity $\urms$ which 
itself is a function of shear and rotation.

\begin{figure}
\centering
\includegraphics[width=1.\columnwidth]{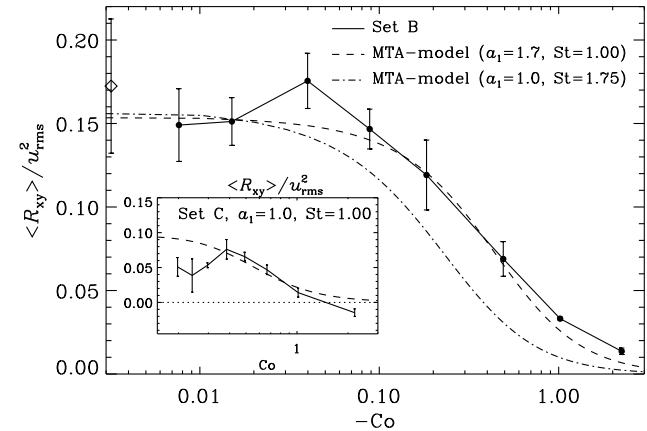}
\caption{Total stress $\rxy$ averaged over the convectively unstable
  layer for the runs in set~B. The dashed
  and dot-dashed curves show results from the minimal $\tau$
  approximation model (Equation~\ref{equ:RxyMTA}). The open symbol on the
  left denotes the stress for run~A4 with no rotation. The inset
  shows the same representation for the runs in set~C.}
   \label{fig:pstressCon2}
\end{figure}

We find that for negative $q$ (Figure~\ref{fig:pstressCon}),
the stress decreases monotonically as rotation is increased. 
For slow rotation, i.e.\ $\Co\gtrsim-0.04$, the differences between 
the runs are not statistically significant, cf.\
Figure~\ref{fig:nut} and Table~\ref{tab:runs}. This is also clear 
from Figure~\ref{fig:pstressCon2} where we plot the volume average 
of the stress over the convectively unstable layer as a function 
of rotation.
For more rapid rotation, we
interpret the decrease of the stress as the non-diffusive 
contribution due to the
$\Lambda$-effect which is likely a good approximation for slow
rotation (see below).
The situation is less clear in the case of positive $q$, see
Figure~\ref{fig:pstressCop} and the inset of Figure~\ref{fig:pstressCon2}. 
A contributing factor is that we cannot use
arbitrarily small $\Omega$ in the positive $q$ regime because cases with
$q>2$ are Rayleigh unstable. 
However, in the runs that can be done
the stress is somewhat decreased in comparison to runs with
corresponding $|\Co|$ and negative $q$. 
For the most rapidly rotating cases in set~C the sign of the stress
changes, which is not observed in set~B.
The sign change suggests that the $\Lambda$-effect dominates over the
turbulent viscosity in the rapid rotation regime.
Similar asymmetry between the regimes of positive and negative $q$
have also been reported e.g.\ by Snellman et al.\ (2009) and Korpi et
al.\ (2010).

It is interesting to see whether simple analytical models can
reproduce the simulation results.
For example, in the presence of weak shear and rotation, the minimal 
$\tau$-approximation applied to a homogeneous system with no 
convection gives (see Appendix~A.2 of Snellman et al.\ 2009) 
\begin{equation}
\langle \rxy \rangle = a_1\,\St\,\frac{-(\Co + \Sh) \, \langle\rxx^{(0)}\rangle+\Co \, \langle\ryy^{(0)}\rangle}{1+4\, \Co \, \St^2 (\Co+\Sh)},\label{equ:RxyMTA}
\end{equation}
where angular brackets denote volume averaging and the superscript 0
refers to a non-rotating and non-shearing reference state. 
A similar result was obtained earlier for arbitrary shear and rotation 
by Narayan et al.\ (1994)
with a conceptually different model
where individual eddies were treated as particles that scatter off 
each other. We note that neither of these models is directly 
applicable to the present system, although treatment of convection 
can be introduced into the models (e.g.\ Kumar et al.\ 1995). 
However, our purpose here is not to perform a detailed comparison 
of the closure models with simulations but rather to attain a 
broad understanding of the system.
In Figure~\ref{fig:pstressCon2}, we compare the numerical results with
the analytical estimate given in Equation~(\ref{equ:RxyMTA}),
keeping the Strouhal number as a free
parameter. Furthermore, we have introduced a scaling parameter $a_1$ (=1.7)
in order to improve the fit. 

We find that parameters $\St=1.0$ and $a_1=1.7$ produce a good fit to
the numerical results for the runs in set~B. We have here normalized
our results from Equation~(\ref{equ:RxyMTA}) by the value of $\urms$ from the
non-rotating run~A0, so a scaling factor $a_1$ greater than unity can in
principle be understood to reflect the decrease of $\urms$ as rotation
is increased.
However, this scaling is not essential
since even the unscaled curve shows qualitatively the same behavior.
The fit for the runs in set~C is not as successful although the simple
model coincides with the simulation data for intermediate values of
$\Co$. However, the simulation results fall below the model for
$q>1.25$ and the negative stresses for rapid rotation are not captured
by the model. The latter is in line with the discovery of Snellman et
al.\ (2009) that the validity of the minimal $\tau$-approximation is
limited to the slow rotation regime.
The lack of proper parameterization of thermal convection in our simple
model is another obvious reason for the differences.

\subsection{$\Lambda$-effect due to shear-induced anisotropy}
In the absence of shear, but including rotation parallel or
antiparallel to
gravity, turbulence is statistically axisymmetric and there is no
asymmetry between the turbulence intensities in the two horizontal
directions, i.e.\ $\rxx = \ryy$. This implies that
there is no horizontal $\Lambda$-effect which, to the lowest order, is
proportional to (R\"udiger 1989)
\begin{equation}
R_{xy}^{(\Omega)}\equiv\Lambda_{\rm H} \Omega_0=2\Omega_0 \tau (R_{yy}-R_{xx}).\label{equ:LH}
\end{equation}
Note that the same result is borne out of Equation~(\ref{equ:taulong})
if we assume that all large-scale flows vanish and allow deviations from
axisymmetry, i.e.\ $\rxx\neq\ryy$.
However, when shear is included, the turbulence becomes anisotropic
in the horizontal plane,
enabling the generation of a non-diffusive contribution to the stress
$R_{xy}$ due to rotation, according to Equation~(\ref{equ:LH}).  Such
contributions have earlier
been studied analytically
(Leprovost \& Kim 2007, 2008a,b) and numerically (Snellman et al.\
2009) for isotropically forced homogeneous turbulence.

\begin{figure}
\centering
\includegraphics[width=1.\columnwidth]{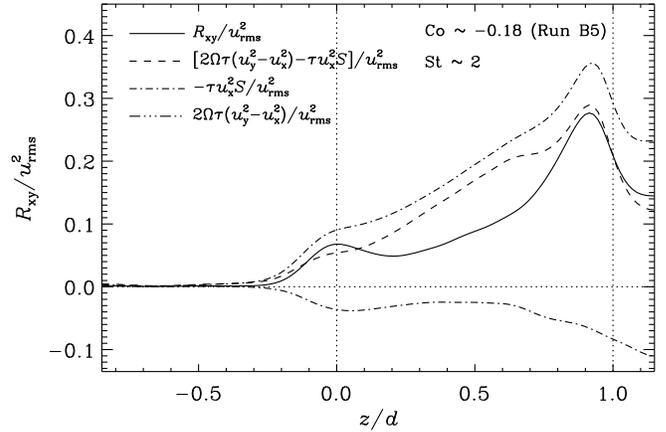}
\caption{Total stress from the simulation (solid line), total stress
  from Equation~(\ref{equ:totstr}) (dashed), diffusive stress from
  Equation~(\ref{equ:mtanut}) (dot-dashed), and non-diffusive
  Equation~(\ref{equ:LH}) (triple-dot-dashed) for run~B5.}
   \label{fig:plammta}
\end{figure}

\begin{figure}
\centering
\includegraphics[width=1.\columnwidth]{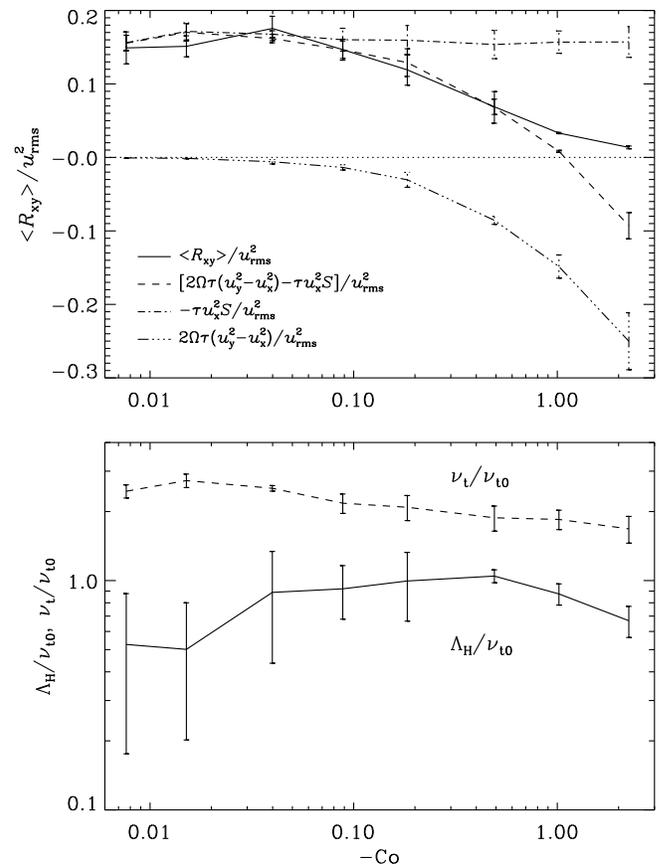}
\caption{Top panel: total stress for runs~B1--B7 (solid line), total
  stress using Equation~(\ref{equ:totstr}) (dashed), and the diffusive
  (dot-dashed) and non-diffusive (triple-dot-dashed) contributions
  according to Equations~(\ref{equ:mtanut}) and (\ref{equ:LH}),
  respectively. Lower panel: coefficients $\Lambda_{\rm H}$ and $\nut$
  according to Equation~(\ref{equ:lamnut}).}
   \label{fig:plambda}
\end{figure}

When both shear and rotation are present it is not possible to
separate the diffusive from the non-diffusive contribution directly.
Furthermore, using the diffusive stress from a purely shearing run to
extract the non-diffusive one from a run with both the effects is also
problematic due to the relatively large errors in the data (cf.\
Figure~\ref{fig:pstressCon2}) which can lead to spurious results. The
large errors in the purely shearing runs can possibly be explained by a
subcritical vorticity dynamo (cf.\ Section~\ref{OnlyShear}).
However, if rotation is slow, we can use the simple analytical results
of Equations~(\ref{equ:mtanut}) and (\ref{equ:LH}) to express the total
stress as
\begin{equation}
\rxy = 2\Omega \tau (\ryy-\rxx) - \tau \rxx S.\label{equ:totstr}
\end{equation}
On the other hand, we can express the stress in terms of the
$\Lambda$-effect and turbulent viscosity by
\begin{equation}
\rxy = \Lambda_{\rm H}\Omega - \nut S,
\end{equation}
where we now have
\begin{equation}
\Lambda_{\rm H}= 2\tau (\ryy-\rxx),\,\, \nut = \tau \rxx,\label{equ:lamnut}
\end{equation}
and where we again treat $\tau$ as a fitting parameter. 
Figure~\ref{fig:plammta}
shows an example from run~B5.
We find that, when $\tau$ corresponding to $\St\approx2$ is used, the
total stress is in broad agreement with Equation~(\ref{equ:totstr}). 
In addition to the weaker negative diffusive contribution,
corresponding to the turbulent viscosity, we find a non-diffusive
part of the opposite sign.  
Reasonably good fits can be obtained for
runs with $|\Co|<1$ with a $\tau$ corresponding to $\St\approx2$. For
more rapidly rotating cases the representation
Equation~(\ref{equ:totstr}) breaks down. Furthermore, in the rapid rotation
regime the relevant time scale is the rotation period rather than the
turnover time.  For the purposes of visualization, and without
altering the qualitative character of the results, we consider here
volume averages over the convectively unstable layer.  Results for
runs~B1--B7 are shown in Figure~\ref{fig:plambda}. We use a fixed
$\tau=8\sqrt{d/g}$ which corresponds to $\St\approx1.5\ldots2.1$ in
these runs. We find that the total stress is fairly well reproduced
for runs where $|\Co|$ is below unity. Furthermore, the diffusive
contribution stays almost constant as a function of $\Co$, whereas the
non-diffusive part is close to zero for $|\Co|<0.1$. The turbulent
viscosity, as obtained from Equation~(\ref{equ:lamnut}), shows a weak
declining trend as a function of rotation, see the lower panel of
Figure~\ref{fig:plambda}. The coefficient $\Lambda_{\rm H}$ has values
in the range $(0.5$--$1)\nutz$ for slow rotation. The error estimates
increase towards slow rotation which is consistent with the fact that
the non-diffusive stress is small at low Coriolis numbers.  These
results suggest that the $\Lambda$-effect is non-zero when the
anisotropy of the turbulence is induced by the shear flow with a
roughly rotation independent coefficients $\Lambda_{\rm H}$. However,
as our method breaks down when $|\Co|\gtrsim1$, we cannot study
quenching behavior of $\nut$ and the $\Lambda$-effect for rapid
rotation.

\begin{figure}
\centering
\includegraphics[width=1.\columnwidth]{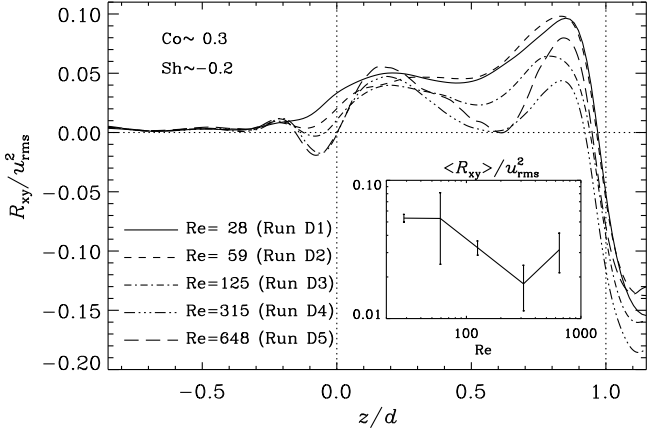}
\caption{Horizontally averaged stress component $\rxy$ from set~D with
  varying Reynolds number and $\Co\approx0.3$ and $\Sh=-0.2$. The
  inset shows the stress averaged over the convectively unstable layer
  as a function of the Reynolds number.}
   \label{fig:pstressRe}
\end{figure}

\subsection{Dependence on Reynolds number}
\label{DependenceReynolds}
In set~D we vary the Reynolds number, keeping shear and rotation
fixed. 
Here we choose $q=1.5$, corresponding to Keplerian shear.
Again, the values of $\Co$ and $\Sh$ are not exactly constant
due to the varying $\urms$; see Table~\ref{tab:kepruns}.
Figure~\ref{fig:pstressRe} shows the results for the horizontally
averaged stress from the runs in set~D. We find that for
relatively weak shear and rotation, the stress is positive in the
convectively unstable layer --- in apparent contradiction with some
earlier results (Stone \& Balbus 1996; Cabot 1996) but in accordance
with the recent results of Lesur \& Ogilvie (2010). We discuss this
issue in the next section in detail.

We also find that the vertical distribution and magnitude of the
stress are not much affected when the Reynolds number is increased
from 28 to 648, see the inset of Figure~\ref{fig:pstressRe} where $\rxy$
averaged over the convectively unstable layer and time is shown. We
note that in addition to the Reynolds number, the Rayleigh number in
this set changes by a factor of 20. However, we have kept the heat
conduction, $K$, and thus the input energy flux, constant so that
$\urms$ varies by only 10\% within set~D. Had we kept $\Pra=1$,
the energy input at the lower boundary would have also decreased by a
factor of 20. This would have resulted in a much lower $\urms$ and thus
proportionally greater values of $\Co$ and $\Sh$. This would have
likely produced a very different trend as a function of $\Ra$ because
$\Co$ and $\Sh$ are considered as the relevant dimensionless 
parameters for the
Reynolds stress (see, e.g., Snellman et al.\ 2009).

In a recent study, Lesur \& Ogilvie (2010) presented results from
Boussinesq convection in an otherwise similar shearing box setup as
ours. They report that the stress changes sign from negative to
positive in the Keplerian case for strong shear when the Rayleigh
number is increased sufficiently. This appears to be in contradiction
with our results regarding the dependence on Rayleigh number.
However, in their setup the Richardson number (Ri), defined as the
negative of the ratio of the squared Brunt--V\"ais\"al\"a frequency
and the squared shear rate, is less than unity. This indicates that
their models are in the shear-dominated regime whereas in our case
only one simulation (run~E8) falls in this regime. The low--Ri case
at high Rayleigh numbers certainly merits further study.

\begin{figure}
\centering
\includegraphics[width=1.\columnwidth]{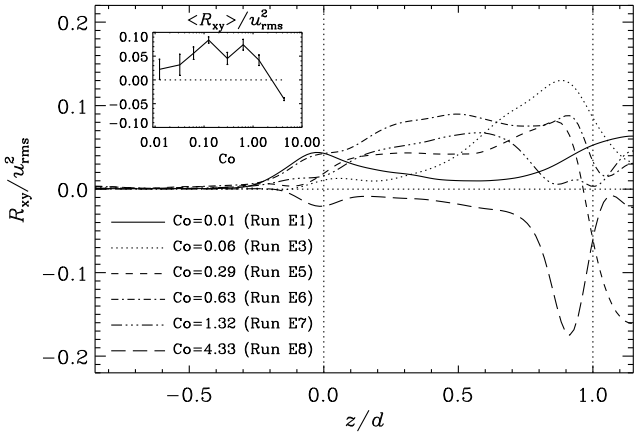}
\caption{Vertical profiles of $R_{\rm xy}$ for several values of $\Co$
  using data for set~E.
  The inset shows $R_{\rm xy}$ averaged over the convectively unstable layer
  as a function of $\Co$.}
   \label{fig:pkepler}
\end{figure}

\begin{figure*}
\centering
\includegraphics[width=1.\textwidth]{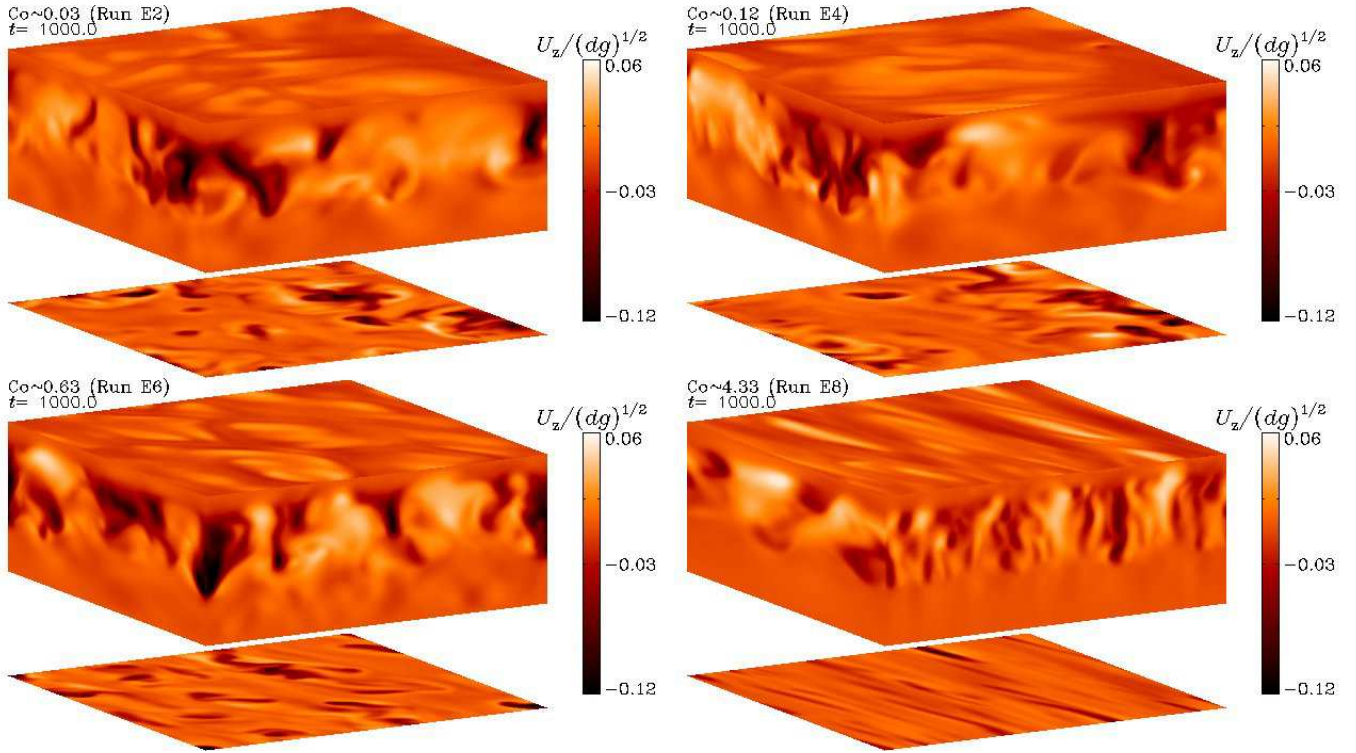}
\caption{Effect of increasing rotation and shear rates on the visual
appearance of the vertical velocity $U_{\rm z}$ (runs~E2, E4, E6 and E8).
All runs are for Keplerian shear, i.e.\ $q=3/2$ and hence $\Sh/\Co=-3/4$,
the resolution is $128^3$ mesh points and the Reynolds number varies
between 20 and 30.}
   \label{fig:boxes}
\end{figure*}

\begin{figure}[t!]
\centering
\includegraphics[width=1.\columnwidth]{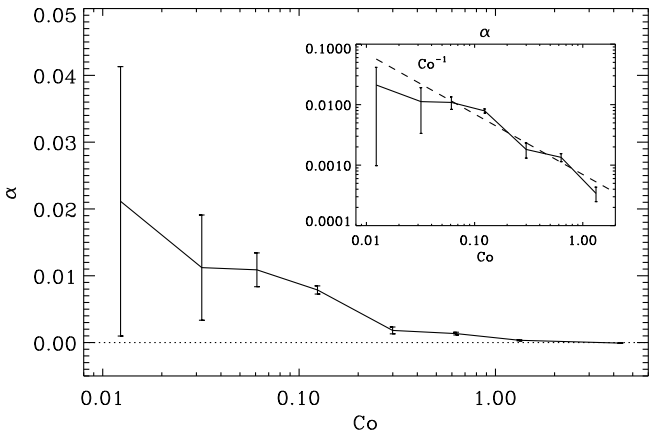}
\caption{Viscosity parameter $\alpha$ according to
  Equation~(\ref{equ:alpha}) for the runs in set~E. The inset show the same
  in log-log scales. The dashed line shows a scaling inversely
  proportional to the Coriolis number for reference.}
   \label{fig:alpha}
\end{figure}

\subsection{Relation to accretion disk theory}
In their paper, Stone \& Balbus (1996) performed a numerical simulation
of convection in a local accretion disk model
and found that the total stress is small and
on average directed inward. This numerical result based on one
simulation and insight drawn from analytical arguments led them to
conclude that convection cannot account for the outward angular
momentum transport required in astrophysical accretion disks.
However, numerical simulations of isotropically forced homogeneous
turbulence under the influence of shear and rotation indicate that
the total stress can change sign as a function of $\Co$ when $q$ is
fixed (Snellman et al.\ 2009). In their study, Snellman et al.\ (2009)
found that, for slow rotation, the stress is positive and changes sign
near $\Co=1$. For rapid rotation ($\Co\approx10$) the stress
appears to drop close to zero. In the context of mean-field
hydrodynamics, this can be understood as quenching of the
$\Lambda$-effect and turbulent viscosity due to shear and rotation,
i.e.\ $\Lambda_{\rm H}=\Lambda_{\rm H}(\Omega,S)$ and
$\nut=\nut(\Omega,S)$.

In an effort to study whether these results carry over to convection
we perform a set of runs where we keep $q=1.5$ fixed and vary the values
of $\Co$ and $\Sh$. Our results for the horizontally averaged stress
are shown in Figure~\ref{fig:pkepler}. For slow rotation and weak shear
$\rxy$ is positive with a maximum value of roughly 10 per cent of the
mean square velocity. As we increase the rotation, the stress decreases and
changes sign for $\Co\approx2$. For $\Co=4.33$, $\rxy$ is negative 
everywhere in the convectively unstable layer.
This is when the flow pattern has become markedly anisotropic,
as can be seen from visualizations of $U_{\rm z}$ on (or near) the
periphery of the computational domain (Figure~\ref{fig:boxes}).
Indeed, the flow pattern becomes rather narrow in the cross-stream
direction, while  being roughly unchanged in the streamwise direction.
In the rapid rotation regime the non-axisymmetric structures tend to
disappear and promote a two-dimensional flow structure that leads to
inward transport (Cabot 1996), which is also visible from our results.

Let us now ask under what conditions can we expect outward angular momentum
transport in accretion disks, i.e., when can one expect the Coriolis number
to be below unity?
In disks we have $\Omega_0 H=\cs$, where $H$ is the scale height and
$\cs$ is the sound speed.
Inserting this into our definition of $\Co$ we obtain from Equation~(\ref{Codef})
\begin{equation}
\Co = \frac{2\Omega_0}{\urms \kef}
=\frac{2\Omega_0 H}{\urms}\,\frac{1}{\kef H}
={2\over\kef H}\,\Ma_{\rm t}^{-1},
\end{equation}
where we have defined the turbulent Mach number $\Ma_{\rm t}=\urms/\cs$.
In our simulations, $\kef H$ is of order unity, which would suggest that
we can expect $\Co<1$ only for supersonic turbulence.

A more refined estimate can be obtained by invoking the Shakura--Sunyaev
$\alpha$-parameter (Shakura \& Sunyaev 1973), which is introduced as a
parameterization of the turbulent viscosity $\nut$ via
\begin{equation}
\nut = \alpha \cs H. \label{equ:SS}
\end{equation}
We can use this parameterization together with Equation~(\ref{nutz})
to eliminate $\urms$ in favor of $\alpha$ under the assumption that
$\nut\approx\nutz$.
This leads to
\begin{equation}
\Co=\frac{2\cs}{3\nut\kef^2H}
=\frac{2}{3\alpha(\kef H)^2}.\label{equ:coalp}
\end{equation}
In our simulations we have $H\approx 0.62d$ at the base of the 
convectively unstable layer.
Together with our estimate $\kef=2\pi/d$ this yields $\kef H\approx4$,
and therefore $\Co\approx(24\alpha)^{-1}$.
We can therefore expect $\Co<1$ for accretion disks where the turbulence
is sufficiently vigorous so that $\alpha\gtrsim0.04$.

We now use our simulations to estimate $\alpha$.
In disks, the rate of strain is proportional to $q\Omega_0$, so
the total turbulent stress is given by
\begin{equation}
T_{xy} = \nut q \Omega_0 , \label{equ:Txy}
\end{equation}
where, in the absence of any other stresses such as from magnetic fields,
the total stress per unit mass is given by
\begin{equation}
T_{xy} = R_{xy}\equiv\mean{u_xu_y}.\label{equ:totalstress}
\end{equation}
Combining Equations~(\ref{equ:SS}) and (\ref{equ:Txy}) gives
\begin{equation}
\alpha=\frac{T_{xy}}{q \Omega_0 \cs H}.\label{equ:alpha}
\end{equation}
We compute $T_{xy}$ and hence $\alpha$ as volume averages over
the convectively unstable region.
For the normalization factor, we take conservative estimates of $\cs$
and $H$ from the bottom of the convectively unstable layer.
We find that for slow rotation, i.e.\ $\Co<0.2$, $\alpha$ is roughly
constant with a value of the order of 0.01 (see
Figure~\ref{fig:alpha}). For more rapid rotation $\alpha$ decreases and
eventually changes sign. The points in the range $\Co\approx
0.06\ldots 1$ are consistent with a scaling inversely proportional to
the Coriolis number which is also suggested by Equation~(\ref{equ:coalp}).
We find that for $\Co\approx1$ we have $\alpha\approx4\cdot10^{-4}$
which is two orders of magnitude smaller than the estimate derived
above.

Another problem facing the suggestion that convection might drive the
angular momentum transport in accretion disks is that without an
internal heat source in the disk, convection is not self-sustained
(cf.\ SB96). However, many disks are likely to be susceptible to the
magnetorotational instability which can extract energy from the shear flow
and ultimately deposit it as thermal energy in the disk. If the
material in the disk is sufficiently thick optically, the pile-up of
energy from the magnetorotational instability could render the
vertical stratification of the disk convectively unstable.

As alluded to in the introduction,
our aim is not to claim that convection is solely responsible for the
outward angular momentum transport in accretion disks but to show that,
given the right conditions, convection can contribute to outward
angular momentum transport.

\section{Conclusions}
\label{sec:conc}
The present results have shown that hydrodynamic convection is
able to transport angular momentum both inward and outward,
depending essentially on the value of the Coriolis number, 
in accordance with earlier results from homogeneous isotropically
forced turbulence (Snellman et al.\ 2009).
This underlines the importance of considering comprehensive
parameter surveys and not to rely on demonstrative results
from individual case studies.
For given value of the Coriolis number, the stress is found to be
relatively independent of the value of the Rayleigh number
(Section~\ref{DependenceReynolds}).
By varying shear and rotation rates separately, we have been
able to quantify the relative importance of diffusive and
non-diffusive contributions to the Reynolds stress tensor.
In agreement with earlier work, it turns out that the turbulent
kinematic viscosity is of the order of the mixing length estimate
and has roughly the same value as the turbulent
magnetic diffusivity found earlier for similar runs
(K\"apyl\"a et al.\ 2009b).
In other words, the turbulent magnetic Prandtl number is around 
unity, again in accordance with results from simpler systems (e.g.\ 
Yousef et al.\ 2003).

The other important turbulent transport mechanism in rotating
turbulent bodies is the $\Lambda$ effect.
Although the importance of this effect is well recognized in
solar and stellar physics (e.g.\ R\"udiger \& Hollerbach 2004), 
it is not normally considered in the
context of accretion disks.
In the present paper we have been able to quantify its importance
for a range of Coriolis numbers by
means of a simple analytical model making use of the minimal 
$\tau$-approximation.
For slow rotation the coefficient $\Lambda_{\rm H}$ is of the order of
$\nutz$ and independent of the Coriolis number.
However, once the Coriolis number exceeds a value around unity,
our method of separating the turbulent viscosity and the
$\Lambda$-effect breaks down,
which
reinforces the need for a truly independent determination not only
of diffusive and nondiffusive contributions to the Reynolds stress,
but also of all the components of the full stress tensor.

\acknowledgements
  The computations were performed on the facilities hosted by CSC - IT
  Center for Science Ltd. in Espoo, Finland, who are administered by
  the Finnish Ministry of Education.
  Financial support from the Academy
  of Finland grants No.\ 136XYZ, 140970 (PJK) and 112020 (MJK)
  are acknowledged. 
  The authors acknowledge the hospitality of NORDITA during their 
  visits.

\vspace{1cm} \noindent {\small \emph{$ $Id: paper.tex,v 1.143 2010-07-02 08:27:08 pkapyla Exp $ $}

\end{document}